 \documentclass[preprint,aps,showpacs]{revtex4}

\usepackage{graphicx}
 
\begin{document}
 \newcommand{\Qed}{\rule{2.5mm}{3mm}}
 \newcommand{\balpha}{\mbox{\boldmath {$\alpha$}}}
 \draft 
%
%
\title{A new understanding of fermion masses from the unified theory of spins and charges 
}
\author{ A. Bor\v stnik Bra\v ci\v c, \\
Educational Faculty, University of Ljubljana,
 Kardeljeva plo\v s\v cad 17, 1000 Ljubljana,\\
 M. Breskvar,  D. Lukman 
and\\
N.S. Manko\v c Bor\v stnik\\
Department of Physics, University of 
Ljubljana, Jadranska 19, 1000 Ljubljana,\\
}
\date{\today}
\begin{abstract} 
In this letter we try to answer those of the open questions 
of the Standard model which concern 
the appearance of families, mass protection mechanism and the Yukawa couplings - by using the 
approach (proposed by one of 
us\cite{norma92,norma93,norma95,holgernorma00,norma01,
Portoroz03,pikanorma05,matjazdragannorma06}), which suggests a new way  
beyond the Standard model.  
The approach has in the starting action for fermions, which carry in $d(=1 +13)-$dimensional space
only the spin (two kinds of the spin) and interact with only spin connection and vielbein 
fields, the term 
manifesting as a mass term in $d=1+3$. (After making several 
approximations and assumptions) we connect free parameters of 
the approach with the experimental data 
and investigate a possibility that the fourth family appears at low enough energies to be 
observable in the new generation of accelerators. 
\end{abstract}


\maketitle 

\noindent
{\em Introduction:} 
The Standard model of the electroweak and strong interactions leaves  unanswered many 
open questions, among which are also the questions   
about the origin of the families, the Yukawa couplings of quarks and leptons and the 
corresponding Higgs mechanism and the weak scale.  
Understanding the mechanism for generating families, their masses and mixing matrices 
might be one of the most promising ways to 
physics beyond the Standard model. 
The approach, unifying spins and charges\cite{norma92,norma93,
norma95,
holgernorma00,norma01,
Portoroz03,pikanorma05,matjazdragannorma06}, might by   
offering a new way of describing families and mass matrices, give an explanation 
about the origin of the Yukawa couplings and show a way beyond the Standard Model. 
It was demonstrated in the refs.\cite{norma01,
Portoroz03,pikanorma05} that a left handed $SO(1,13)$ 
Weyl spinor multiplet includes, if the representation is analyzed 
in terms of the subgroups $SO(1,3)$, $SU(2)$, $SU(3)$ and the sum of the two $U(1)$'s,  all 
the spinors of the Standard model - that is the left handed $SU(2)$ doublets and the right 
handed  $SU(2)$ singlets of (with the group  $SU(3)$ charged) quarks and  (chargeless) leptons -  
answering the question where does originate the connection between the weak charge 
and the handedness (which concerns only the spin in $d=1+3$). \\
The approach assumes  two kinds of  the spin connection  fields - the gauge fields of 
the two kinds of the Clifford algebra objects - and a vielbein field in 
$d=(1+13)-$dimensional space, 
which might\footnote{
The approach seems to have, like all the Kaluza-Klein-like theories, a very 
serious disadvantage, namely that there might not exist  any massless, 
mass protected spinors, which are, after the break of symmetries, 
chirally coupled to the desired (Kaluza-Klein) gauge fields\cite{witten}. 
This would mean that there are no observable spinors at low energies. 
We have tried hard to find an  example\cite{holgernorma05}, a toy model, 
 which  gives a hope to Kaluza-Klein-like 
theories by demonstrating that a kind of a break of symmetries does lead to 
massless, mass protected spinors, 
chirally coupled to the Kaluza-Klein gauge fields, observable at low energies.
}  manifest - after some  
appropriate compactifications (or some other kind of making the rest of $d-4$ space 
unobservable at low energies) - in the four dimensional 
space as all the gauge fields of the known charges, as well as the Yukawa couplings, determining 
mass matrices of families of quarks and leptons and accordingly their masses and 
mixing matrices. 
This letter is a 
short review of the two papers\cite{pikanorma05,matjazdragannorma06}, which  analyze 
how do terms, which lead to masses of quarks 
and leptons, appear in the approach unifying spins and charges as a part of the 
(vacuum expectation values of) spin 
connection and vielbein fields. No Higgs is needed in this approach to ''dress'' 
right handed spinors with the weak charge, since the terms of the 
starting Lagrangean, which include $\gamma^0\gamma^s,$ with $s=7,8,$ do the job of a Higgs 
field of the Standard model. \\

\noindent
{\em Two kinds of the Clifford algebra objects:} We assume two kinds of the Clifford 
algebra objects defining two kinds of the generators of the 
Lorentz algebra\cite{norma92,norma93,holgernorma02,technique03}.  
One kind are the ordinary Dirac $\gamma^a$ operators defining the generators of the Poincar\' e algebra 
for spinors $S^{ab}$ ($S^{ab}= \frac{1}{4} (\gamma^a \gamma^b - \gamma^b \gamma^a)$). 
The second kind\footnote{The operators $\tilde{\gamma}^a$  are introduced 
formally as operating on any Clifford 
algebra object $B$ 
from the left hand side, but they also can be expressed in terms of the  ordinary $\gamma^a$ as  
operating from the right hand side as follows 
$\tilde{\gamma}^a B : = i(-)^{n_B} B \gamma^a,$
with $(-)^{n_B} =  +1$ or $-1$, when the object $B$ has a Clifford even or odd character, respectively.}
of the Clifford objects $\tilde{\gamma}^a$ commutes with 
$\gamma^a$ ($\{\tilde{\gamma}^a,\tilde{\gamma}^a\}_+=0$) 
and defines accordingly 
$\tilde{S}^{ab}$ ($\tilde{S}^{ab} = \frac{1}{2} (\tilde{\gamma}^a \tilde{\gamma}^b-
\tilde{\gamma}^b \tilde{\gamma}^a)$), with $\{\tilde{S}^{ab}, S^{cd}\}_-=0$. They are responsible for 
the generation of families.\\ 
We define a basis of spinor representations as  
eigen states of the chosen Cartan subalgebra of the Lorentz algebra $SO(1,13)$,  with  the operators 
$S^{ab}$ and $\tilde{S}^{ab}$ in the two Cartan subalgebra sets, with the same indices in both cases.
When introducing the notation\cite{holgernorma02,technique03,pikanorma05}
\begin{eqnarray}
\stackrel{ab}{(\pm i)}: &=& \frac{1}{2}(\gamma^a \mp  \gamma^b),  \quad 
\stackrel{ab}{[\pm i]}: = \frac{1}{2}(1 \pm \gamma^a \gamma^b), \;{\rm for} \; \eta^{aa} \eta^{bb} =-1,
\nonumber\\
\stackrel{ab}{(\pm )}: &= &\frac{1}{2}(\gamma^a \pm i \gamma^b),  \quad 
\stackrel{ab}{[\pm ]}: = \frac{1}{2}(1 \pm i\gamma^a \gamma^b), \;{\rm for} \; \eta^{aa} \eta^{bb} =1,
\label{eigensab}
\end{eqnarray}
it can be shown that  the above binomials are all ''eigen vectors''  of  the generators $S^{ab}$, 
as well as of  $\tilde{S}^{ab}$
\begin{eqnarray}
S^{ab} \stackrel{ab}{(k)} &=&  \frac{k}{2} \stackrel{ab}{(k)}, \quad 
S^{ab} \stackrel{ab}{[k]}  =  \frac{k}{2} \stackrel{ab}{[k]}, \nonumber\\
\tilde{S}^{ab} \stackrel{ab}{(k)}  &= & \frac{k}{2} \stackrel{ab}{(k)},  \quad 
\tilde{S}^{ab} \stackrel{ab}{[k]}   =   - \frac{k}{2} \stackrel{ab}{[k]}.
\label{eigensabev}
\end{eqnarray}
Defining $ 
\stackrel{ab}{\tilde{(k)}}= \frac{1}{2} (\tilde{\gamma}^a + 
\frac{\eta^{aa}}{ik} \tilde{\gamma}^b ) $ we find the relations 
\begin{eqnarray}
         \gamma^a \stackrel{ab}{(k)}&=&\eta^{aa}\stackrel{ab}{[-k]},\quad 
          \gamma^b \stackrel{ab}{(k)}= -ik \stackrel{ab}{[-k]}, \nonumber\\
         \gamma^a \stackrel{ab}{[k]}&=& \stackrel{ab}{(-k)},\quad \quad \quad
          \gamma^b \stackrel{ab}{[k]}= -ik \eta^{aa} \stackrel{ab}{(-k)},\nonumber\\
\tilde{\gamma^a} \stackrel{ab}{(k)} &=& - i\eta^{aa}\stackrel{ab}{[k]},\quad
\tilde{\gamma^b} \stackrel{ab}{(k)} =  - k \stackrel{ab}{[k]}, \nonumber\\
\tilde{\gamma^a} \stackrel{ab}{[k]} &=&  \;\;i\stackrel{ab}{(k)},\quad \quad \quad
\tilde{\gamma^b} \stackrel{ab}{[k]} =  -k \eta^{aa} \stackrel{ab}{(k)},\nonumber\\
\stackrel{ab}{(k)}\stackrel{ab}{(k)}& =& 0, \quad \quad \stackrel{ab}{(k)}\stackrel{ab}{(-k)}
= \eta^{aa}  \stackrel{ab}{[k]}, \quad 
\stackrel{ab}{[k]}\stackrel{ab}{[k]} =  \stackrel{ab}{[k]}, \quad \quad
\stackrel{ab}{[k]}\stackrel{ab}{[-k]}= 0,  \nonumber\\
\stackrel{ab}{(k)}\stackrel{ab}{[k]}& =& 0,\quad \quad \quad \stackrel{ab}{[k]}\stackrel{ab}{(k)}
=  \stackrel{ab}{(k)}, \quad \quad 
\stackrel{ab}{(k)}\stackrel{ab}{[-k]} =  \stackrel{ab}{(k)},
\quad \quad \stackrel{ab}{[k]}\stackrel{ab}{(-k)} =0,  \nonumber\\
\stackrel{ab}{\tilde{(k)}} \stackrel{ab}{(k)}& =& 0, 
\quad \quad \stackrel{ab}{\tilde{(-k)}} \stackrel{ab}{(k)}
= -i \eta^{aa}  \stackrel{ab}{[k]}, 
\quad \stackrel{ab}{\tilde{(k)}}\stackrel{ab}{[k]}= i \stackrel{ab}{(k)},\quad
\stackrel{ab}{\tilde{(k)}} \stackrel{ab}{[-k]} = 0. 
\label{graphbinomsfamilies}
\end{eqnarray}\\

\noindent
{\em  A Weyl spinor in $d= (1+13)$ manifesting at ''physical energies'' families of 
quarks and leptons:} We assume a  left handed Weyl spinor  in $(1+13)$-dimensional space. 
We make a choice of  $d/2=7$ Cartan subalgebra members in both sectors as follows: 
$S^{03}, S^{12}, S^{56}, S^{78}, S^{9\;10}, S^{11\;12}, S^{13\;14}$ and 
$\tilde{S}^{03}, \tilde{S}^{12}, \tilde{S}^{56}, \tilde{S}^{78}, \tilde{S}^{9\;10}, 
\tilde{S}^{11\;12}, \tilde{S}^{13\;14}$.
We present in Table I all the quarks of one particular colour (the right handed weak 
chargeless $u_R,d_R$ and the left handed weak charged $u_L, d_L$, with the colour 
$(1/2,1/(2\sqrt{3}))$ in 
the  Standard model notation). They all are  members of one $SO(1,7)$ multiplet.  
\begin{table}
\begin{center}
\begin{tabular}{|r|c||c||c|c||c|c|c||c|c|c||r|r|}
\hline
i&$$&$|^a\psi_i>$&$\Gamma^{(1,3)}$&$ S^{12}$&$\Gamma^{(4)}$&
$\tau^{13}$&$\tau^{21}$&$\tau^{33}$&$\tau^{38}$&$\tau^{41}$&$Y$&$Y'$\\
\hline\hline
&& ${\rm Octet},\;\Gamma^{(1,7)} =1,\;\Gamma^{(6)} = -1,$&&&&&&&&&& \\
&& ${\rm of \; quarks}$&&&&&&&&&&\\
\hline\hline
1&$u_{R}^{c1}$&$\stackrel{03}{(+i)}\stackrel{12}{(+)}|\stackrel{56}{(+)}\stackrel{78}{(+)}
||\stackrel{9 \;10}{(+)}\stackrel{11\;12}{(-)}\stackrel{13\;14}{(-)}$
&1&1/2&1&0&1/2&1/2&$1/(2\sqrt{3})$&1/6&2/3&-1/3\\
\hline 
2&$u_{R}^{c1}$&$\stackrel{03}{[-i]}\stackrel{12}{[-]}|\stackrel{56}{(+)}\stackrel{78}{(+)}
||\stackrel{9 \;10}{(+)}\stackrel{11\;12}{(-)}\stackrel{13\;14}{(-)}$
&1&-1/2&1&0&1/2&1/2&$1/(2\sqrt{3})$&1/6&2/3&-1/3\\
\hline
3&$d_{R}^{c1}$&$\stackrel{03}{(+i)}\stackrel{12}{(+)}|\stackrel{56}{[-]}\stackrel{78}{[-]}
||\stackrel{9 \;10}{(+)}\stackrel{11\;12}{(-)}\stackrel{13\;14}{(-)}$
&1&1/2&1&0&-1/2&1/2&$1/(2\sqrt{3})$&1/6&-1/3&2/3\\
\hline 
4&$d_{R}^{c1}$&$\stackrel{03}{[-i]}\stackrel{12}{[-]}|\stackrel{56}{[-]}\stackrel{78}{[-]}
||\stackrel{9 \;10}{(+)}\stackrel{11\;12}{(-)}\stackrel{13\;14}{(-)}$
&1&-1/2&1&0&-1/2&1/2&$1/(2\sqrt{3})$&1/6&-1/3&2/3\\
\hline
5&$d_{L}^{c1}$&$\stackrel{03}{[-i]}\stackrel{12}{(+)}|\stackrel{56}{[-]}\stackrel{78}{(+)}
||\stackrel{9 \;10}{(+)}\stackrel{11\;12}{(-)}\stackrel{13\;14}{(-)}$
&-1&1/2&-1&-1/2&0&1/2&$1/(2\sqrt{3})$&1/6&1/6&1/6\\
\hline
6&$d_{L}^{c1}$&$\stackrel{03}{(+i)}\stackrel{12}{[-]}|\stackrel{56}{[-]}\stackrel{78}{(+)}
||\stackrel{9 \;10}{(+)}\stackrel{11\;12}{(-)}\stackrel{13\;14}{(-)}$
&-1&-1/2&-1&-1/2&0&1/2&$1/(2\sqrt{3})$&1/6&1/6&1/6\\
\hline
7&$u_{L}^{c1}$&$\stackrel{03}{[-i]}\stackrel{12}{(+)}|\stackrel{56}{(+)}\stackrel{78}{[-]}
||\stackrel{9 \;10}{(+)}\stackrel{11\;12}{(-)}\stackrel{13\;14}{(-)}$
&-1&1/2&-1&1/2&0&1/2&$1/(2\sqrt{3})$&1/6&1/6&1/6\\
\hline
8&$u_{L}^{c1}$&$\stackrel{03}{(+i)}\stackrel{12}{[-]}|\stackrel{56}{(+)}\stackrel{78}{[-]}
||\stackrel{9 \;10}{(+)}\stackrel{11\;12}{(-)}\stackrel{13\;14}{(-)}$
&-1&-1/2&-1&1/2&0&1/2&$1/(2\sqrt{3})$&1/6&1/6&1/6\\
\hline\hline
\end{tabular}
\end{center}
\caption{\label{TableI}%
The 8-plet of quarks - the members of $SO(1,7)$ subgroup, 
belonging to one Weyl left 
handed ($\Gamma^{(1,13)} = -1 = \Gamma^{(1,7)} \times \Gamma^{(6)}$) spinor representation of 
$SO(1,13)$. It contains the left handed weak charged quarks and the right handed weak 
chargeless quarks of a particular colour ($(1/2,1/(2\sqrt{3}))$). Here  $\Gamma^{(1,3)}$ 
defines the handedness in $(1+3)$ space, 
$ S^{12}$ defines the ordinary spin (which can also be read directly from the basic vector, 
since, in particular, $S^{12} \stackrel{12}{(+)} = \frac{1}{2} \stackrel{12}{(+)}$), 
$\tau^{13}$ defines the third component of the weak charge, 
$\tau^{21}$ defines the $U(1)$ charge, $\tau^{33}$ and 
$\tau^{38}$ define the colour charge and $\tau^{41}$ another $U(1)$ charge, 
which together with the first $U(1)$ charge  defines $Y = \tau^{21}+ \tau^{41}$ 
and $Y'=-\tau^{21}+ \tau^{41}$. Leptons differ 
from quarks only with respect to the part which concerns the indices $9,\cdots,14$. A singlet 
in the colour, which belongs to the same Weyl, must be of the kind 
$\cdots|\cdots||\stackrel{9 \;10}{(+)}\stackrel{11\;12}{[+]}\stackrel{13\;14}{[+]}$ (one finds that 
$\nu_{R}$ looks like $\stackrel{03}{(+i)}\stackrel{12}{(+)}|\stackrel{56}{(+)}\stackrel{78}{(+)}
||\stackrel{9 \;10}{(+)}\stackrel{11\;12}{[+]}\stackrel{13\;14}{[+]}$)
The reader can find the whole Weyl representation in the ref.\cite{Portoroz03}. }
\end{table}
Looking at the first row of Table I, for example,  and using Eq.(\ref{graphbinomsfamilies}) (saying that 
$\gamma^0 \stackrel{03}{(+i)}=\stackrel{03}{[-i]}$, $\stackrel{78}{(-)}\stackrel{78}{(+)}=-
\stackrel{78}{[-]}$) one sees that $\gamma^0 \stackrel{78}{(-)}$ transforms a right handed 
weak chargeless $u_R$ quark of a particular colour into a left handed weak charged $u_L$ quark 
of the same colour and the spin, presented in the seventh row. One also can notice that 
the generator $\tilde{S}^{07}=\frac{i}{2} \tilde{\gamma}^{0} \tilde{\gamma}^7 $ transforms the 
expression of the first row on Table I 
into with respect to 
$S^{ab}$ an equivalent state ($\stackrel{03}{[+i]}\stackrel{12}{(+)}|\stackrel{56}{(+)}\stackrel{78}{[+]}
||\stackrel{9 \;10}{(+)}\stackrel{11\;12}{(-)}\stackrel{13\;14}{(-)}$, which is again a 
right handed weak chargeless $u_R$ quark of the same colour and spin
), which differs in properties  
from the $u_R$ of the first row only with respect to $\tilde{S}^{ab}$. \\
Assuming that a kind of breaking symmetries\cite{holgernorma05} makes a starting Weyl spinor 
in $d=1+13$ to manifest  after a break into $SO(1,7)\times SU(3)\times U(1)$ 
as massless spinors - one  $SU(3)$ triplet    and one   $SU(3)$ singlet 
 - each of them a member of an  $SO(1,7)$ octet\footnote{Antiquarks and antileptons appear 
 in the second quantized procedure 
as their charged conjugate partners .}, there must be accordingly also 
eight families since one can easily notice that each member of the octet on Table II carries  
two indices: the index of the row and the family index. Namely,  any $\tilde{S}^{ab}; a,b \in 0,1..,8$, 
not belonging to the Cartan subalgebra, transforms any member of the starting family, represented 
on Table I, into another family, with the same spin and charges. \\

\noindent
{\em  A Weyl spinor in $d= (1+13)$ in the two kinds of spin connection fields:}
A spinor carries 
only the spin (no charges) and interacts accordingly with only the gauge gravitational fields 
- with  vielbeins and two kinds of spin connection fields -  the gauge fields of $p^a, S^{ab}$ 
and $\tilde{S}^{ab}$, respectively\cite{norma92,norma93,
holgernorma00,norma01,
Portoroz03,pikanorma05}. 
One kind is the ordinary gauge field (gauging the Poincar\' e symmetry in $d=1+13$). 
The contribution of this field to the mass matrices manifests in only the diagonal terms -  
connecting the right handed weak chargeless quarks or leptons to the left handed weak charged 
partners within one family of spinors. 
The second kind of gauge fields is in our approach responsible for families of spinors and 
couplings among families of spinors - contributing to diagonal matrix elements as well - and    
might explain the appearance of families of quarks and leptons and the Yukawa couplings of the 
Standard model. 
We write the action\cite{pikanorma05} for a Weyl (massless) spinor  
in $d(=1+13)$ - dimensional space as follows\footnote{Latin indices  
$a,b,..,m,n,..,s,t,..$ denote a tangent space (a flat index),
while Greek indices $\alpha, \beta,..,\mu, \nu,.. \sigma,\tau ..$ denote an Einstein 
index (a curved index). Letters  from the beginning of both the alphabets
indicate a general index ($a,b,c,..$   and $\alpha, \beta, \gamma,.. $ ), 
from the middle of both the alphabets   
the observed dimensions $0,1,2,3$ ($m,n,..$ and $\mu,\nu,..$), indices from the bottom of 
the alphabets
indicate the compactified dimensions ($s,t,..$ and $\sigma,\tau,..$). We assume the signature 
$\eta^{ab} =
diag\{1,-1,-1,\cdots,-1\}$.
}
\begin{eqnarray}
S &=& \int \; d^dx \; {\mathcal L}
\nonumber\\
{\mathcal L} &=& \frac{1}{2} (E\bar{\psi}\gamma^a p_{0a} \psi) + h.c. = \frac{1}{2} 
(E\bar{\psi} \gamma^a f^{\alpha}{}_a p_{0\alpha}\psi) + h.c. \nonumber\\
&=& \bar{\psi}\gamma^{m} (p_{m}- \sum_{A,i}\; g^{A}\tau^{Ai} A^{Ai}_{m}) \psi  + \; 
 \sum_{s=7,8}\; \bar{\psi} \gamma^{s} p_{0s} \; \psi + \; {\rm the \;rest},
\nonumber\\
p_{0\alpha} &=& p_{\alpha} - \frac{1}{2}S^{ab} \omega_{ab\alpha} - \frac{1}{2}\tilde{S}^{ab} 
\tilde{\omega}_{ab\alpha}.
\label{lagrange}
\end{eqnarray}
Here $f^{\alpha}{}_a$ are  vielbeins (inverted to the gauge field of the generators of translations  
$e^{a}{}_{\alpha}$, $e^{a}{}_{\alpha} f^{\alpha}{}_{b} = \delta^{a}_{b}$,
$e^{a}{}_{\alpha} f^{\beta}{}_{a} = \delta_{\alpha}{}^{\beta}$),
with $E = \det(e^{a}{}_{\alpha})$, while  
$\omega_{ab\alpha}$ and $\tilde{\omega}_{ab\alpha} $ are the two kinds of the spin connection fields, 
the gauge 
fields of $S^{ab}$ and $\tilde{S}^{ab}$, respectively. 
Index $A$ determines the charge groups ($SU(3), SU(2)$ and the two $U(1)$'s), index $i$ determines
the generators within one charge group. $\tau^{Ai}$ denote the generators of the charge groups 
$\tau^{Ai} = \sum_{s,t} \;c^{Ai}{ }_{st} \; S^{st}, \;
\{\tau^{Ai}, \tau^{Bj}\}_- = i \delta^{AB} f^{Aijk} \tau^{Ak},$
 with $s,t \in 5,6,..,14$, while $A^{Ai}_{m}, m=0,1,2,3,$ 
denote the corresponding
gauge fields (expressible in terms of $\omega_{st m}$). (We assume that no terms like 
$\sum_{A,i}\; \tilde{g}^{A} \tilde{\tau}^{Ai} \tilde{A}^{Ai}_{m}$ manifest at "low energy region", 
a justification for such an assumption will be discussed in a  separate paper.)
The subgroups  and accordingly the coefficients $c^{Ai}{ }_{st}$ are chosen so that the 
gauge fields in the "physical" region agree with the
known gauge fields, while ''the rest'' in Eq.(\ref{lagrange}) is assumed to be small. 
The term $\sum_{s=7,8}\; \bar{\psi} \gamma^{s} p_{0s} \; \psi $ in Eq.(\ref{lagrange}) 
manifests as  the  Yukawa couplings of  the Standard model and we rewrite it as ${\mathcal L}_{Y}$ in 
the following way 
\begin{eqnarray}
{\mathcal L}_{Y} &=& \psi^{\dagger} \gamma^0 \;  
\{ \stackrel{78}{(+)}  (\sum_{y=Y,Y'}\; y A^{y}_{+} + 
\frac{-1}{2}\sum_{(ab)}\; \tilde{S}^{ab} \tilde{\omega}_{ab+}))
+  \quad 
  \stackrel{78}{(-)}  (\sum_{y=Y,Y'}\;y  A^{y}_{-} + 
  \frac{-1}{2}\sum_{(ab)}\; \tilde{S}^{ab} \tilde{\omega}_{ab-})
+ \nonumber\\
 & & \stackrel{78}{(+)} \sum_{\{(ac)(bd) \},k,l} \; \stackrel{ac}{\tilde{(k)}} \stackrel{bd}{\tilde{(l)}}
\tilde{{A}}^{kl}_{+}((ac),(bd)) \;\;+  
 \stackrel{78}{(-)} \sum_{\{(ac)(bd) \},k,l} \; \stackrel{ac}{\tilde{(k)}}\stackrel{bd}{\tilde{(l)}}
\tilde{{A}}^{kl}_{-}((ac),(bd))\}\psi,
\label{yukawa4tilde0}
\end{eqnarray}
with $k,l=\pm 1,$ if $\eta^{aa}\eta^{bb}=1$ and 
$ \pm i,$ if $\eta^{aa}\eta^{bb}=-1$, while $Y$ and $Y'$ 
are the two superpositions of the two $U(1)$ subgroups of the groups $SO(6)$ and $SO(1,7)$ as  defined 
in refs.\cite{pikanorma05,matjazdragannorma06}. We rewrote  
 $\sum_{(a,b) } -\frac{1}{2}\; \stackrel{78}{(\pm)}\tilde{S}^{ab} \tilde{\omega}_{ab\pm} =
\sum_{(cd)} \; - \frac{1}{2}\; \stackrel{78}{(\pm)}\tilde{S}^{ab} \tilde{\omega}_{ab\pm} \; +  
\sum_{(ac),(bd), \;  k,l}\stackrel{78}{(\pm)}\stackrel{ac}{\tilde{(k)}}\stackrel{bd}{\tilde{(l)}} 
\; \tilde{A}^{kl}_{\pm} ((ac),(bd))$, where the pair $(a,b)$ in the first sum runs over all 
the  indices, with $ a,b = 0,\dots, 8$,  while the  pairs in the second and the third 
sum $(cd),(ac),(bd)$ 
denote only the Cartan subalgebra pairs. Accordingly all the pairs $(ab)$ in Eq.(\ref{yukawa4tilde0}) 
run only over Cartan subalgebra pairs. 
The Yukawa part of the starting Lagrangean (Eq.(\ref{lagrange})) has the diagonal terms, that is the 
terms manifesting the Yukawa couplings within each family, and the off diagonal terms, determining the 
Yukawa couplings among families, as we shall demonstrate bellow. \\
One notices that in Eq.(\ref{yukawa4tilde0}) only the part with the factor 
$\gamma^0 \stackrel{78}{(+)}$  contributes to  the mass matrix of the $d-$quarks and the electrons, 
while the part with the factor $\gamma^0 \stackrel{78}{(-)}$  contributes to only 
the mass matrix of the $u-$quarks and the neutrinos. 
The first four sums of Eq.(\ref{yukawa4tilde0}) contribute to 
only diagonal terms of either the $d-$quarks and the electrons or the $u-$quarks and neutrinos. 
Terms with $(Y,Y')$ distinguish among the $u_R-$quarks ($Y=2/3,Y'=-1/3$), the $d_R-$quarks 
($Y=-1/3,Y'=2/3$), the left handed quarks ($Y=1/6,Y'=1/6$), the $\nu_R$ ($Y=0,Y'=-1$), 
the $e_R$ ($Y=-1,Y'=0$) and the left handed leptons ($Y=-1/2,Y'=-1/2$). The terms 
with $\tilde{S}^{ab}$ do not distinguish among the members of one family, they only differ 
in values for different families. 
The last two sums in Eq.(\ref{yukawa4tilde0}) contribute to the non diagonal matrix elements of either 
the $u-$quarks and the neutrinos (the term with the factor $\gamma^0 \stackrel{78}{(-)}$) or 
to the $d-$quarks and the electrons  (the term with the factor $\gamma^0 \stackrel{78}{(+)}$).\\

\noindent
{\em Breaks of symmetries and observable properties:} We made the assumption that a 
break of symmetries  leads  from one Weyl representation in $d=1+13$ to four massless octets 
(the representation of $SO(1,7)$) with the charges $SU(3),Y$ and $Y'$ as presented above, leaving  
us with eight  equivalent representations - that is with eight families. In order to be 
in agreement with what we observe, we must break further the symmetry of the octet in 
the charge sector. Namely, looking at Table I and recognizing\cite{pikanorma05} that 
$Q= \tau^{33}+ Y= S^{56} + \tau^{41}$  must appear as a conserved 
quantity representing the electromagnetic charge, we assume that no terms of the types  
$S^{5a} \omega_{5a\pm}$ and $S^{6a} \omega_{6a\pm}$, with $a \ne 5,6$ may appear in our 
${\mathcal L}_{Y}$. Assuming that the break influences both sectors - $S^{ab}$ and  $\tilde{S}^{ab}$ -  
in a similar way, we let also all the terms 
$\tilde{S}^{5a} \tilde{\omega}_{5a\pm}$ and $\tilde{S}^{6a} \tilde{\omega}_{6a\pm}$, 
with $a \ne 5,6$, contribute nothing, 
which means that we assume the break of $SO(1,7)$ into $SO(1,5)\times U(1)$, 
which further means that 
eight families decouple entirely into two times four families. We shall not try to justify better 
this  assumption in this letter. Instead, we shall study, what can we learn  
from our approach (after all 
these assumptions about the appropriate breaks of the starting symmetry). 
The first row on Table I, representing the $u_R$ quark,  then appears in the following four 
families (while all the other members of a particular family 
follow from the one for $u_R$  by the application of $S^{ab}; a,b \in \{0,8\}$ and 
equivalently for the quarks of other two colours and  for the colourless leptons) 
\begin{eqnarray}
I.\;& & \stackrel{03}{(+i)} \stackrel{12}{(+)} |\stackrel{56}{(+)} \stackrel{78}{(+)}||...\nonumber\\ 
II.\;& &\stackrel{03}{[+i]} \stackrel{12}{[+]} |\stackrel{56}{(+)} \stackrel{78}{(+)}||... \nonumber\\
III.& & \stackrel{03}{[+i]} \stackrel{12}{(+)} |\stackrel{56}{(+)} \stackrel{78}{[+]}||... \nonumber\\
IV. & & \stackrel{03}{(+i)} \stackrel{12}{[+]} |\stackrel{56}{(+)} \stackrel{78}{[+]}||... \;.
\label{fourfamilies}
\end{eqnarray}
Obviously, starting from one Weyl in $d$ we only can have an even number of families. We 
have measured up to now three families. If we further break the symmetry, like $SO(1,5)$ into 
$SU(2)\times SU(2)\times U(1)$, by letting, for example,  all the terms $\tilde{S}^{7a}\;
\tilde{\omega}_{7a\pm}$ and $\tilde{S}^{8a}\;
\tilde{\omega}_{8a\pm},a\ne7,8$,  
contribute nothing, we shall end up with twice two completely decoupled families. If we 
break instead $SO(1,5)$ into $SU(3)\times U(1)$, one family decouples from the rest three. 
The experimental data seems at least for quarks to be closer to an assumption of an approximate break 
of $SO(1,5)$ to $SU(2)\times SU(2)\times U(1)$, since the first two families 
are much lighter than the third and 
also the quark mixing matrix seems not to disagree with such an assumption. We shall accordingly 
assume the following approximate shape of any of the mass matrices for the quarks and the leptons 
\begin{equation}\label{deggen}
          \left(\begin{array}{cc}
                  A   &  B\nonumber\\
                  B   &  C=A+k B \nonumber\\
                    \end{array}
                \right), 
\end{equation}
with $A,B,C$ which are two by two matrices and where k is a constant. All these matrices 
ought to be calculable in terms of the fields  
in Eq.(\ref{yukawa4tilde0}). We shall assume 
that after the suggested breaks of symmetries the fields appearing in Eq.(\ref{yukawa4tilde0}) 
take a vacuum expectation values and (after the integration over all but $1+3$ dimensions) manifest 
in $d=1+3$ as parameters. We shall further simplify our mass matrices by
assuming that all the matrix elements are real and that the matrices are symmetric. 
We then end up  with the mass 
matrices\cite{pikanorma05,matjazdragannorma06} expressed in terms of the parameters $a_\alpha, 
\tilde{\omega}_{abc} $ as one finds on Table \ref{TableII}. \\
\begin{table}
\begin{center}
\begin{tabular}{|r||c|c|c|c|}
\hline
$\alpha$&$ I_{R} $&$ II_{R} $&$ III_{R} $&$ IV_{R}$\\
\hline\hline
$I_{L}$   & $ a_{\alpha}  $ & $ \frac{1}{2}(\tilde{\omega}_{327\alpha} +\tilde{\omega}_{018\alpha}) $ & $  
\frac{1}{2}(\tilde{\omega}_{387\alpha} +\tilde{\omega}_{078\alpha}) $  &
$  \frac{1}{2}\tilde{\omega}_{187\alpha} $ \\
\hline
$II_{L}$  & $ \frac{1}{2}(\tilde{\omega}_{327\alpha} +\tilde{\omega}_{018\alpha}) $
& $a_{\alpha} +  (\tilde{\omega}_{127\alpha} - \tilde{\omega}_{038\alpha})  $ & 
$\frac{1}{2}\tilde{\omega}_{187\alpha} $&$\frac{1}{2}(\tilde{\omega}_{387\alpha} 
- \tilde{\omega}_{078\alpha})$ \\
\hline 
$III_{L}$ & $ \frac{1}{2}(\tilde{\omega}_{387\alpha} +\tilde{\omega}_{078\alpha})$ & $
\tilde{\omega}_{187\alpha} $ & $ \frac{k_{\alpha}}{2}(\tilde{\omega}_{387\alpha} +  
\tilde{\omega}_{078\alpha})$ &
$ \frac{k_{\alpha}}{2} \tilde{\omega}_{187\alpha}  $ \\
\hline 
$IV_{L}$  & $ \frac{1}{2}\tilde{\omega}_{187\alpha}$ & 
$\frac{1}{2}(\tilde{\omega}_{387\alpha} -\tilde{\omega}_{078\alpha}) $ & 
$ \frac{k_{\alpha}}{2} \tilde{\omega}_{187\alpha}$ & $ 
\frac{k_{\alpha}}{2}  (\tilde{\omega}_{387\alpha} - \tilde{\omega}_{078\alpha}) $ \\
\hline\hline
\end{tabular}
\end{center}
\caption{\label{TableII}%
The mass matrix of four families of  quarks and leptons, obtained within the approach 
unifying spins and charges under the assumptions\cite{pikanorma05} that an appropriate break of the 
starting symmetry  to $SO(1,7)\times SU(3)\times U(1)$ leads to massless quarks and 
leptons, while in further breaks the electromagnetic charge is conserved - and equivalently in 
the $\tilde{\omega}_{abc}$ sector - that an approximate break occurs  further from $SO(1,5)$ to 
$SU(2)\times SU(2)\times U(1)$ leading to the symmetry of Eq.(\ref{deggen})
(ref.\cite{matjazdragannorma06}), that 
the mass matrices are real and symmetric and that evaluation can be done 
on ''a tree level''. }
\end{table}

\noindent
The parameters on Table~\ref{TableII} are related as follows 
\begin{eqnarray}
\label{b}
k_u = -k_d, \; k_{\nu} = -k_e,&\;&\tilde{\omega}_{387u} = -b_{387u}\;\tilde{\omega}_{387 d},\nonumber\\ 
\tilde{\omega}_{127u} = \; b_{127u}\; \tilde{\omega}_{127 d},&& \quad
\tilde{\omega}_{078u}  =  \; b_{127u}\; \tilde{\omega}_{078 d}, \nonumber\\
\tilde{\omega}_{018u} = -b_{018u}\;\tilde{\omega}_{018 d},&& \quad
\tilde{\omega}_{187u}  = \; b_{018u}\;\tilde{\omega}_{187 d},
\end{eqnarray}
and similarly for leptons, when $u$ is replaced by $\nu$ and $d$ by $e$. 
It is the Lagrange density ${\mathcal L}_Y$ of Eq.(\ref{yukawa4tilde0}) which determines the relations 
among the parameters - after making all the discussed assumptions (for which we only have a kind 
of explanation, no justification since we have not yet treated  breaking of symmetries 
and nonperturbative effects). If we assume that breaks of symmetries and nonperturbative 
effects influence all the quarks and the leptons in the same way, factors $b_{abc\alpha}$ are 
$1$ and $\tilde{\omega}_{abc\alpha}$ are the same for quarks and leptons.
Since we do not know how breaks of symmetries and 
nonperturbative effects  influence the parameters of mass matrices, 
we  let $b_{abc\alpha}$ as well as $\tilde{\omega}_{abc \alpha}$ be different 
for quarks and for leptons. We study what the fit to the experimental data can tell about 
the parameters\cite{pikanorma05,matjazdragannorma06}.
The parameter $a_{\alpha}$ distinguishes among the members of 
one family, since it is a sum of the diagonal contributions of 
$-\frac{1}{2}\tilde{S}^{ab} \tilde{\omega}_{ab\pm} $ and the 
contributions of $\tau^y A^{y}_{7,8},$ $y=Y,Y'$. \\


\noindent
{\em Connecting free parameters of the approach with the experimental data:} It is easy to see that 
any $4\times 4$ matrix of the form of Eq.(\ref{deggen}) is diagonizable with three (rather than  
with six) angles and that the angles of rotations are related\cite{matjazdragannorma06} 
as follows 
\begin{eqnarray}
\tan (\varphi_{\alpha}-\varphi_{\beta}) = \pm \frac{k_{\alpha}}{2},\; 
({\rm or} \: \pm \frac{2}{k_{\alpha}}),\nonumber\\
\tan ({}^{a,b}\varphi_{\alpha}-{}^{a,b}\varphi_{\beta}) = \pm \frac{{}^{a,b}\eta_{\alpha}}{2},\; 
({\rm or} \: \pm \frac{2}{{}^{a,b}\eta_{\alpha}}), 
\label{angles}
\end{eqnarray}
with $\alpha =u,\nu$, $\beta = d,e$, ${}^{a}\eta_{\alpha} = \frac{\tilde{\omega}_{127\alpha} + 
\sqrt{1+(k_{\alpha}/2)^2}\; \tilde{\omega}_{078 \alpha}}{ \tilde{\omega}_{018 \alpha} - 
 \sqrt{1 + (k_{\alpha}/2)^2}\;\tilde{\omega}_{187\alpha}}= - {}^{a}\eta_{\beta}$ and 
  ${}^{b}\eta_{\alpha} = \frac{\tilde{\omega}_{127\alpha} - 
 \sqrt{1+(k_{\alpha}/2)^2}\; \tilde{\omega}_{078 \alpha}}{ \tilde{\omega}_{018 \alpha} + 
 \sqrt{1 + (k_{\alpha}/2)^2}\;\tilde{\omega}_{187\alpha}}= - {}^{b}\eta_{\beta}$, while 
 the index $a$ determines the first two by two mass matrix and $b$ determines the second 
 two by two mass matrix after the first step  diagonalization of $4 \times 4$ mass matrix 
 (determined by the angle $\varphi_{\alpha}$) is performed. One further finds that
 $\varphi_{\alpha} = \frac{\pi}{2}  -\varphi_{\beta}, 
 \varphi = \varphi_{\alpha} -\varphi_{\beta},$ and 
 ${}^{a,b}\varphi_{\alpha} = \frac{\pi}{2}  - {}^{a,b}\varphi_{\beta}, 
  {}^{a,b}\varphi = {}^{a,b}\varphi_{\alpha} -{}^{a,b}\varphi_{\beta}.$ \\
The experimental mixing matrices for quarks and for leptons are not in contradiction 
with our (rough) assumption that they are symmetric and real (up to charge parity 
symmetry breaking, which in this letter is not  treated) and it also turns out that both can
within the experimental accuracy be fitted with three angles\cite{expckm,expmixleptons}. 
Due to the refs.\cite{okun,okunmaltoni,okunbulatov} the fourth family is 
experimentally not excluded. 
We proceed by fitting the experimental data of the two mixing matrices and of the masses 
of the known three families of quarks and leptons by the requirement that the ratios among the 
$\tilde{\omega}_{abc}$ (that is $b_{abc\alpha}$) are as close to the rational numbers 
of small integers as possible, having in mind that a kind of breaking symmetries or nonperturbative  
effects  or both reflect  the influence of charges. 
We use the Monte-Carlo simulation program. What we 
obtained\cite{matjazdragannorma06} is represented in Table \ref{TableIII} and 
Table \ref{TableIV} and in Eqs.(\ref{resultmasses},\ref{resultckm},\ref{resultckmleptons}). 
We see on Table \ref{TableIV} that all the parameters $b_{abc\alpha}$ are either very close to 
$\frac{1}{2}$ or to $2$.
\begin{table}
\centering
\begin{tabular}{|c||c|c|c|c|}
\hline 
&$u$&$d$&$\nu$&$e$\tabularnewline
\hline
\hline 
$k$       & -0.085 &  0.085&-1.254&  1.254\tabularnewline
\hline 
${}^a\eta$& -0.229 &  0.229& 1.584& -1.584\tabularnewline
\hline
${}^b\eta$&  0.420 & -0.440& -0.162& 0.162\tabularnewline
\hline
\end{tabular}\\
\caption{\label{TableIII}%
The Monte-Carlo fit to the experimental data\cite{expckm,expmixleptons} 
for the three parameters $k$, ${}^a\eta$ and   ${}^b\eta$ determining the mixing matrices 
for  the four families of quarks and leptons.}
\end{table}
\begin{table}
\begin{tabular}{|c||c|c|c||c|c|c|}
\hline 
&$u$&$d$&$b_u$&$\nu$&$e$&$b_{\nu}$\tabularnewline
\hline
\hline 
$|\tilde{\omega}_{018}|$& 21205& 42547& 0.498&    10729& 21343 &0.503\tabularnewline
\hline 
$|\tilde{\omega}_{078}|$& 49536& 101042& 0.490&   31846& 63201 &0.504\tabularnewline
\hline 
$|\tilde{\omega}_{127}|$& 50700& 101239& 0.501&   37489& 74461 &0.503\tabularnewline
\hline 
$|\tilde{\omega}_{187}|$& 20930& 42485& 0.493&    9113&  18075 &0.505\tabularnewline
\hline 
$|\tilde{\omega}_{387}|$& 230055& 114042& 2.017&  33124& 67229 &0.493\tabularnewline
\hline
$a^{a}$&94174& 6237& &   1149&1142 &\tabularnewline
\hline
\end{tabular}\\
\caption{\label{TableIV}%
Values for the parameters $\tilde{\omega}_{abc}$ (entering into 
the mass matrices for the $u-$quarks, the $d-$quarks, the neutrinos and the 
electrons suggested by the approach - Table \ref{TableII},Eq.\ref{b}) 
as following after the Monte-Carlo fit  
relating the parameters and the experimental data.}
\end{table}
\noindent
We find for the masses of the four families the values
\begin{eqnarray}
\label{resultmasses}
m_{u_i}/GeV &=& (0.0034, 1.15, 176.5, 285.2),\nonumber\\
m_{d_i}/GeV &=& (0.0046, 0.11, 4.4, 224.0), \nonumber\\
m_{\nu_i}/GeV &=& ( 1\; 10^{-12}, 1 \; 10^{-11}, 5 \; 10^{-11},  84.0 ),\nonumber\\
m_{e_i}/GeV &=& (0.0005,0.106,1.8, 169.2).
\end{eqnarray}
\noindent
For the quark mixing matrix we found 
\begin{eqnarray}
\label{resultckm}
 \left(\begin{array}{cccc}
 0.974 & 0.223 & 0.004 & 0.042\\
 0.223 & 0.974 & 0.042 & 0.004\\
 0.004 & 0.042 & 0.921 & 0.387\\
 0.042 & 0.004 & 0.387 & 0.921\\
 \end{array}
                \right)
\end{eqnarray}
and for the leptons we found 
\begin{eqnarray}
\label{resultckmleptons}
 \left(\begin{array}{cccc}
 0.697 & 0.486 & 0.177 & 0.497\\
 0.486 & 0.697 & 0.497 & 0.177\\
 0.177 & 0.497 & 0.817 & 0.234\\
0.497  & 0.177 & 0.234 & 0.817\\ 
\end{array}
                \right).
\end{eqnarray}\\

\noindent
{\em Concluding remarks:} 
We have studied in this letters what the approach unifying spins and charges - offering 
a new way beyond the Standard model of the electroweak and colour interactions - can say about 
the origin of families and the Yukawa couplings. The letter is a short review of the 
two papers\cite{pikanorma05,matjazdragannorma06}. We have started 
%
 with the action for 
a Weyl spinor in $d=1+13$, which carries two kinds of spins and no charges and interacts with only 
the gravity 
We have assumed  breaks of symmetries 
which end up with  a nonzero part of the starting Lagrange density ($ 
\psi^{\dagger} \gamma^0 \stackrel{78}{(\pm)} p_{0\pm}\psi$ 
(Eq.(\ref{yukawa4tilde0})) manifesting as Yukawa couplings - the mass term  
transforming a right handed weak chargeless quark or lepton into the corresponding 
left handed weak charged one - in $(1+3)-$dimensional space. No Higgs is needed. \\ 
The approach predicts an even number 
of families with the mass matrices determined with the two kinds of gauge fields. One kind 
determines only the matrix elements within a family and distinguishes among the members 
of a family, the second kind determines diagonal and of diagonal matrix elements. We made several 
assumptions and approximations to come to simple expressions, which enable us 
to fit parameters of the approach with the existing experimental data.  
Not knowing how possible breaks of symmetries influence the starting mass matrix with 
$10$ free parameters ($5 \; \tilde{\omega}_{abc}$ and  $4$ times 
$ a_{\alpha}$ with one relation among 
$a_{\alpha}$ and $2$ times 
$k_{\alpha}$) we allow $22$ parameters ($4$ times $5 \; \tilde{\omega}_{abc\alpha} + 4 $ 
times $ a_{\alpha}$ plus $2$ times $k_{\alpha}$, which are related by  
$4$ times ${}^{a,b}\eta_{\alpha}$), which we further relate by the requirement 
that the ratios $b_{abc\alpha}$ (Eq.(\ref{b}))
are as close to small rational numbers as possible ($2$ times $5$ requirements),  
 to be fitted using a Monte-Carlo procedure with $2\times 3$ angles 
 and $4\times 3$ masses within the experimental accuracy. 
\\
 Our rough estimate predicts that masses of the fourth family might be 
measurable with new accelerators. Correspondingly also mixing matrices for quarks and 
leptons are predicted. All these predictions must be taken with caution. \\
We treat quarks and leptons equivalently, assuming that breaking symmetries causes equal effects on 
quarks and leptons. (Assuming, for example, that quarks feel the break
of $SO(1,5) $ approximately into $SU(2)\times SU(2)\times U(1)$, while leptons follow 
the  approximate break to $SU(3)\times U(1)$, would change the prediction for leptons.) 
We do not take a possibility of the Majorana spinors into account.\\
If our approach unifying spins and charges is showing the right way 
beyond the Standard model, it will answer the open questions about the origin 
of families and the Yukawa couplings. To try to predict the properties without all  
the assumptions and approximations we did and without fitting parameters of 
the approach with the experimental data (and accordingly also predicting the weak scale) 
is a huge project, 
to which this letter (and the refs.\cite{pikanorma05,matjazdragannorma06}) is a small first step. 
In any case this first step might open a new window into understanding the reasons for 
the assumptions of the Standard model. 

\noindent
{\em Acknowledgments:} 
It is a pleasure to thank all the participants of the   workshops entitled "What comes beyond 
the Standard model", taking place  at Bled annually in  July, starting at 1998,  for fruitful 
discussions, in particular H.B. Nielsen.

 \end{document}